\documentclass[twocolumn,dvipsnames]{aastex62}

\usepackage{CJK}
\usepackage{framed,amsmath,bm,booktabs}
\usepackage{float}

\newcommand{\pp}[2]{\frac{\partial#1}{\partial#2}}

\graphicspath{{./}{figures}}

\submitjournal{ApJ}
\shorttitle{Cosmic Ray Mediated Thermal Fronts}
\shortauthors{Zhu, Zweibel, Gnedin}

\begin{document}
\begin{CJK*}{UTF8}{gkai}

\title{Cosmic Ray Mediated Thermal Fronts in the Warm-Hot Circumgalactic Medium}

\correspondingauthor{Hanjue Zhu (朱涵珏)}
\email{hanjuezhu@uchicago.edu}
\author[0000-0003-0861-0922]{Hanjue Zhu (朱涵珏)}
\affiliation{Department of Astronomy \& Astrophysics; 
The University of Chicago; 
Chicago, IL 60637, USA}

\author{Ellen G.\ Zweibel}
\affiliation{Department of Astronomy, University of Wisconsin-Madison, 475 N. Charter Street, Madison, WI 53706, USA}
\affiliation{Department of Physics, University of Wisconsin-Madison, 150 University Avenue, Madison, WI 53706, USA}

\author[0000-0001-5925-4580]{Nickolay Y.\ Gnedin}
\affiliation{Particle Astrophysics Center; 
Fermi National Accelerator Laboratory;
Batavia, IL 60510, USA}
\affiliation{Kavli Institute for Cosmological Physics;
The University of Chicago;
Chicago, IL 60637, USA}
\affiliation{Department of Astronomy \& Astrophysics; 
The University of Chicago; 
Chicago, IL 60637, USA}

\begin{abstract}
We investigate the 1D plane-parallel front connecting the warm ($10^4$ K) and hot ($10^6$ K) phases of the circumgalactic medium (CGM), focusing on the influence of cosmic rays (CRs) in shaping these transition layers. We find that cosmic rays dictate the thermal balance while other fluxes (thermal conduction, radiative cooling, and gas flow) adjust to compensate. We compute column density ratios for selected transition temperature ions and compare them with observational data. While most of our models fail to reproduce the observations, a few are successful, although we make no claims for their uniqueness. Some of the discrepancies may indicate challenges in capturing the profiles in cooler, photoionized regions, as has been suggested for by previous efforts to model thermal transition layers.
\end{abstract}

\section{Introduction}\label{sec:intro}

The circumgalactic medium (CGM), a gaseous halo enveloping galactic disks, acts as both a reservoir for star formation and a repository for galactic outflows and feedback processes. Its multiphase structure, characterized by regions of varying temperatures and densities, retains the imprints of the galaxy's formation and evolutionary history. Investigating the CGM is crucial for understanding the baryon cycle in galaxies, as it regulates the flow of gas between the galaxy and the intergalactic medium, thereby influencing star formation rates and the effectiveness of feedback processes \citep{Tumlinson2017, FG23}.

Recent observational and theoretical studies have significantly advanced our understanding of the CGM. Observations using metal absorption lines, such as those from O VI, C IV, and Mg II, have unveiled the complex multiphase nature of the CGM, highlighting substantial variations in temperature, density, and ionization states across different regions \citep{Prochaska2017,Tumlinson2017}. Discrepancies between observed line ratios and model predictions indicate the necessity for models beyond collisional ionization equilibrium (CIE) and photoionization equilibrium (PIE) to accurately interpret CGM properties. Efforts to address these discrepancies include considering radiative cooling flows that incorporate gas dynamics and self-photoionization, turbulent mixing layers where Kelvin-Helmholtz instabilities create intermediate-temperature regions around clouds, conductive interfaces where cool clouds evaporate and hot gas condenses at the surface due to heat conduction, and ionized gas behind radiative shocks potentially produced by strong galactic winds \citep{Wakker2012, Ji2019, Armillotta2017, Gnat2009}.

An important aspect of the CGM is the influence of cosmic rays (CRs). CRs provide additional pressure support, affect thermal balance, and may drive galactic winds, all of which shape the overall structure and evolution of the CGM. Their ability to propagate over vast distances and recoup their losses by extracting energy from ambient shocks and turbulence allows CRs to impact regions far from their origin, influencing the thermal and dynamical properties of the CGM on a larger scale \citep{Salem2016,Butsky2018}. However, in the absence of observational evidence for cosmic rays in the CGM, we can only speculate upon their influence. And because direct detection of cosmic rays in diffuse environments such as the CGM is challenging, we must resort to observational predictions based on modeling the thermal gas itself.

The transfer of momentum and energy between cosmic rays and thermal gas is mediated by magnetic field structures on a scale of order the cosmic ray gyroradius. These small-scale magnetic structures could be part of an extrinsic turbulent cascade, but can also be driven by the cosmic rays themselves via the so-called streaming instability \citep{Kulsrud1969}. The latter case is known as cosmic ray self confinement, and we will assume throughout this paper that it holds. An interesting consequence of the theory is that it breaks down if the projections of the cosmic ray pressure gradient and thermal plasma density gradient along the background magnetic field oppose each other. In this case - as predicted from analytical arguments by \cite{Skilling1971}  and shown first in a simulation by \cite{Wiener2017}, scattering disappears, the cosmic rays decouple from the gas, and their pressure becomes constant. Such transport bottlenecks should be common in the CGM, with its clumpy density structure, provided the clumps are magnetically connected to the Galactic disk, where most cosmic rays presumably originate.

Bottlenecks were studied in a series of papers of increasing generality in \cite{Wiener2017,Wiener2019,Bustard2021,Tsung2022}. These works showed that the large CR pressure gradient accelerates cool clouds upward, imparting significant momentum and causing changes in gas density and temperature profiles. Additionally, CRs were found to pressurize the cold gas, reducing its density and heating the cloud's surface, leading to broader transition layers and altering ion abundances. These findings underscore the significant role of CRs in shaping the thermal and dynamical properties of the CGM. Due to the disparity in scale between the assumed cloud size and the front thickness, the transition region has only been modeled previously as a 1D structure in a steady state \citep{Wiener2017}. 

The transition layers between the cool and hot phases, characterized by a density contrast typically around a factor of 100 as temperatures shift from $10^4$ K to $10^6$ K, are particularly affected by CRs. Models of these transition layers are crucial not only for explaining observed line ratios but as was shown in \citep{Wiener2017}, may also be indicators of clouds dominated by cosmic ray pressure. This dominance would signify that CRs are a significant component of the galactic structure, affecting both the local environment of the CGM and the broader galactic ecosystem; it would also be evidence for the theory of cosmic ray self confinement.

In this study, we refine the 1D plane-parallel symmetric model presented by \citet{Wiener2017} by incorporating thermal conduction and gas flow in the energy equation in addition to cosmic ray heating and radiative cooling. These additions aim to broaden the range of cosmic ray-mediated front models and provide a more comprehensive understanding of the CGM's structure and dynamics.

The plan of this paper is as follows. In \S\S\ref{ss:review}, we briefly review some well-known front models. In \S\S\ref{ss:eqns} we formulate the problem, with a separate discussion of the boundary conditions in \S\S\ref{sec:ic}. Section \ref{sec:results} gives detailed discussions of two types of front, static (\S\S\ref{sec:results:static_front}) and evaporative (\S\S\ref{sec:results:evaporation}). We show the effects of changing the background magnetic field and cosmic ray pressure in the cloud in \S\ref{sec:background}. In \S\ref{sec:ratios} we compute the column densities of several transition temperature ions across our range of front models and compare their ratios to the ratios derived by \cite{Wakker2012} from absorption line spectroscopy. Section \ref{sec:discussion} summarizes and discusses our main conclusions. Appendix A is a short discussion of condensation fronts. Appendix B shows some results on how the column densities of transition temperature ions depend on domain over which we integrate.

\section{Physical Setup}\label{sec:setup}
\subsection{Review of Front Models}\label{ss:review}
Past work on fronts without cosmic rays provides valuable context for understanding the distinct role of CRs in gas dynamics. \citet{Cowie1977} investigated evaporative fronts in the interstellar medium, where thermal conduction from a surrounding hot medium drives the continuous evaporation of cooler gas clouds. Their model established a steady-state balance between thermal conduction and enthalpy flux as the heating and cooling terms, respectively, with mass being lost from the cool phase to the hot phase. They also demonstrated how the geometry of the system influences the stability and evolution of these fronts. \citet{Inoue2006} later examined phase transition layers connecting two thermal equilibrium phases, a cold neutral medium cooled by C II and a warm neutral medium cooled by Ly$\alpha$, with heating provided by photoelectrons from dust. Because both studies solved for temperature assuming energy balance and heat transfer by classical conduction\footnote{Cowie \& McKee also considered saturated heat flux, which does not follow Fick's Law, but that is immaterial for our purposes.} they were treated as two-point boundary value problems. In \cite{Cowie1977}, the temperature of the hot medium was chosen based on observations and the cloud temperature was approximated by zero. \citet{Inoue2006} set the boundary temperatures to their equilibrium values.

In \citet{Wiener2017}, thermal conduction and enthalpy flux were dropped, leaving collisionless cosmic ray heating supplemented by a base heating rate per particle to balance radiative cooling. Without heat conduction,  the energy equation was reduced to first order. The models were integrated outward for 100 pc, at which point $T$ had typically increased from its chosen value of 10$^4$ K at the cloud boundary to a few 10$^6$ K. Over this temperature range it was not possible to find a reasonable high-temperature thermal equilibrium, so the model was solved as an initial value problem and the temperature was still increasing at the 100 pc cutoff. It was checked \textit{a posteriori} that the fronts found from the model were too thick for heat transport by thermal conduction to play a significant role. We expand on this issue in \S\ref{sec:results}. 

\begin{table*}[htbp]
\centering
\begin{tabular}{ccc}
\toprule
\textbf{Parameter} & \textbf{Fiducial Value} & \textbf{Range} \\
\midrule
$n_{p,0}$ & $0.05$ cm$^{-3}$ & Fixed \\ 
$n_0$ & $0.1$ cm$^{-3}$ & Fixed \\ 
$T_0$ & $10^4$ K & Fixed \\ 
$\kappa_T$ & $5.6 \times 10^{-7} T^{5/2}$ erg s$^{-1}$ K$^{-1}$ cm$^{-1}$ & Fixed \\ 
$B$ & $3 \ \mu$G & $3 - 30 \ \mu$G \\ 
$\Gamma$ & $1 \times 10^{-26}$ erg s$^{-1}$ & Fixed \\ 
$\Lambda(T)$ & $1.1 \times 10^{-21} \times 10^{\Theta(\log(T / 10^5 \mathrm{~K}))}$ erg cm$^3$ s$^{-1}$ & Fixed \\ 
$\vec{v_0}$ & $400$ m s$^{-1}$ & $40 - 400$ m s$^{-1}$ \\ 
$P_{g,0}$ & $1.38 \times 10^{-13}$ erg cm$^{-3}$ & Fixed \\ 
$P_{c,0}$ & $(\alpha-1) P_{g,0}$ & $2 P_{g,0} - 9P_{g,0}$ \\ 
$\alpha$ & 3 & $3 - 10$ \\
\bottomrule
\end{tabular}
\caption{Parameters values adopted in this paper.}
\label{table:params}
\end{table*}

\subsection{Gas and Cosmic Ray Equations}\label{ss:eqns}
The evolution equations governing a thermal gas describe the conservation of mass, momentum, and energy. In 1D, they are:
\begin{equation}\label{eq:gas1}
\frac{\partial \rho}{\partial t} + \frac{\partial (\rho v)}{\partial x} = 0,
\end{equation}
\begin{equation}\label{eq:gas2}
\frac{\partial (\rho v)}{\partial t} + \frac{\partial}{\partial x} (\rho v^2 + P_g + P_c) = 0,
\end{equation}
\begin{equation}\label{eq:gas3}
\begin{split}    
\frac{\partial E_g}{\partial t} + \frac{\partial}{\partial x} \left[(E_g + P_g)v - \kappa_T \frac{\partial T}{\partial x}\right] = \\ -\rho \mathcal{L} - (v+v_A) \frac{\partial P_c}{\partial x},
\end{split}
\end{equation}
where $\rho$ and $v$ represent the gas density and velocity, respectively; $v_A$ is the Alfv\'en velocity $B/\sqrt{4\pi\rho}$; $P_g$ and $E_g$ denote the thermal pressure and total energy density of the gas, with 
\begin{displaymath}
E_g = \frac{P_g}{\gamma_g - 1} + \frac{1}{2} \rho v^2,    
\end{displaymath} 
$\kappa_T$ stands for the thermal conduction coefficient, and $\rho \mathcal{L}$ is the net cooling function. For simplicity, we assume a fully ionized Hydrogen plasma and an ideal gas equation of state, and take $\gamma_g = 5/3$ throughout.

We use a net cooling function $\rho \mathcal{L} \equiv n^2 \Lambda - n \Gamma$. We adopt $\Gamma = 1 \times 10^{-26}$ erg/s, consistent with the value used in \citep{Inoue2006}. Because of the sharp cooling drop-off near $T = 10^4$ K, this choice of $\Gamma$ allows for thermal equilibrium at both $T \sim 10^4$ K and $T \sim 10^6$ K. We note that this value of $\Gamma$ is higher than the lower limit set by Coulomb scattering of cosmic rays in thermal plasma \citep[e.g.][]{Yoast-Hull2013}. For the cooling function, we use
\begin{equation}
\Lambda(T) = 1.1 \times 10^{-21} \times 10^{\Theta\left(\log \left(T / 10^5 \mathrm{~K}\right)\right)} \mathrm{erg~cm}^3 \mathrm{~s}^{-1},
\end{equation}
where
\begin{equation}
\Theta(x) = 0.4 x - 3 + \frac{5.2}{e^{x + 0.08} + e^{-1.5(x + 0.08)}}.
\end{equation}
This analytical fit for the cooling function was first given in \citet{Imada2012} and modified in \citet{Wiener2019}. The fit is designed to match the solar metallicity cooling rates from \citet{Wiersma2009}, where they used the Cloudy photoionization code to calculate cooling functions for gas with varying metallicities, temperatures and densities under a UV/X-ray background at $z=0$.

In the fluid approximation, cosmic rays are governed by the energy equation \citep{Mckenzie1982, Breitschwerdt91}:
\begin{equation}\label{eq:CREnergy}
\pp{P_c}{t}=(\gamma_c - 1)(v+v_A)\pp{P_c}{x}-\pp{F_c}{x}+Q,
\end{equation}
where $Q$ accounts for CR sources and sinks and $F_c$ represents the CR energy flux:
\begin{equation}
F_c=\gamma_cP_c (v+v_A) -\kappa_c\pp{P_c}{x}.
\end{equation}
Here, $\kappa_c$ is the CR diffusion coefficient and we take $\gamma_c\equiv 1+P_c/E_c = 4/3$  throughout the paper.

The influence of cosmic rays is manifested through the cosmic ray pressure ($P_c$) gradient term in the momentum equation and the heating term in the energy equation. As shown in \citet{Wiener13,Zweibel17}, streaming cosmic rays heat gas at the rate of $-v_A \nabla P_c$. At a microscopic level, the heating represents the transfer of cosmic ray energy to Alfv\'en waves due to the streaming instability together with the transfer of wave energy to the thermal background, as must occur in a steady state. The pressure gradient force represents momentum transfer between the cosmic rays and the magnetic field fluctuations, and occurs whether cosmic rays are self confined or confined by extrinsic turbulence.

We seek steady state solutions of Equations~(\ref{eq:gas1}) - (\ref{eq:CREnergy}) under the further simplifying assumptions that cosmic ray diffusion is negligible and $v/v_A\ll 1$. That is, we set all time derivatives to zero, drop $\kappa_c$, and wherever $v+v_A$ appears, we replace it by $v_A$. This leads to the following three conserved quantities:

\begin{equation}\label{eq:conserved}
\begin{aligned}
    \rho v &= \mathrm{constant},\\
    \rho v^2 + P_g + P_c &= \mathrm{constant},\\
    P_c/\rho^{\gamma_c/2} &= \mathrm{constant}.
\end{aligned}
\end{equation}

From equations~\ref{eq:conserved} 
we obtain the following relationship between number density ($n$) and temperature ($T$):
\begin{equation}
\rho v^2 + n k_B T+ P_{c, 0} (\frac{n}{n_0})^{2/3} = \rho_{0} v_0^2 + (\alpha+1) n_0 k_B T_0,
\end{equation}
where the ``0" subscripts denote initial values. This allows us to treat $T$ as a function of $n$. We then recast eqn. (\ref{eq:gas3}) as a second order ordinary differential equation (ODE)  for $n$. We approach this ODE system as an initial value problem to determine the spatial profile of the variables. The boundary values are established within the warm ionized medium, as detailed in Section~\ref{sec:ic}.

\subsection{Initial Value Problem: Boundary Values} \label{sec:ic}

In the warm ionized medium on the left boundary ($x = 0$), we set the proton number density ($n_{p,0}$) at $0.05$ cm$^{-3}$, resulting in a total (electron plus proton) number density ($n_0$) of $0.1$ cm$^{-3}$. The initial temperature ($T_0$) is established at $10^4$ K, leading to an initial gas pressure ($P_{g,0} = n_0 k_B T_0$) of $1.38 \times 10^{-13}$ erg/cm$^3$. These boundary conditions are chosen to represent typical conditions in the warm ionized medium of the CGM. In this environment, the thermal conduction coefficient is given by $\kappa_T = 5.6 \times 10^{-7} T^{5/2}$ erg/s/K/cm \citep{Spitzer1962}.

In our fiducial case, we adopt a constant magnetic field strength $B = 3 \ \mu G$ and assume the field is perpendicular to the plane of the front. This choice aligns with observations of the $\sim \mu G$ magnetic field strength in the CGM \citep{Lan2020,Heesen2023,Bockmann2023}.

For the evaporative solutions, we set the gas velocity at the boundary ($v_0$) to 400 m/s, neglecting the effects of any relative motion between the cloud and the ambient medium. The positive value of the mass flux corresponds to the warm medium evaporating into the hot medium. We also explore the scenario with a static front, where $v = 0$. We show in Appendix~\ref{appendix:condensation} that a condensation front cannot form under these conditions. In our setup, we have $v_A \gg v$, which satisfies the condition for the validity of Equation~\ref{eq:gas3}. In our fiducial case, we assume $P_{c,0} = 2 P_{g,0}$ (a value consistent with the bottleneck formation studies of \cite{Wiener2017,Wiener2019}), but the initial ratio of CR pressure to gas pressure is a parameter we vary; we introduce the parameter $\alpha \equiv P_{c,0} / P_{g,0} + 1$. We list all relevant parameters in Table~\ref{table:params}.

\section{Results}\label{sec:results}

We proceed by solving the energy equation (Equation~(\ref{eq:gas3}), reformulated as a second-order
initial value problem for $n$, imposing two boundary conditions: $n_0 = 0.1$ cm$^{-3}$ and $\frac{dn}{dx}\big|_0$, the value of which is varied to explore the effect of the gradient (which determines the strength of various flux terms) on our outcomes. We defer discussions of how $\frac{{dn}}{{dx}}$ is selected in a natural environment to future works, except for a few remarks in \S\ref{sec:discussion}. In \cite{Wiener2017} the front models are generated by integrating a first order ODE, and $\frac{dn}{dx}\big|_0$ is determined through the polytropic relationship between $P_c$ and $n$, along with the requirement that cosmic ray heating balances radiative cooling.

\subsection{Static Front} \label{sec:results:static_front}

\begin{figure}[htb!]
\centering
\includegraphics[width=0.9\columnwidth]{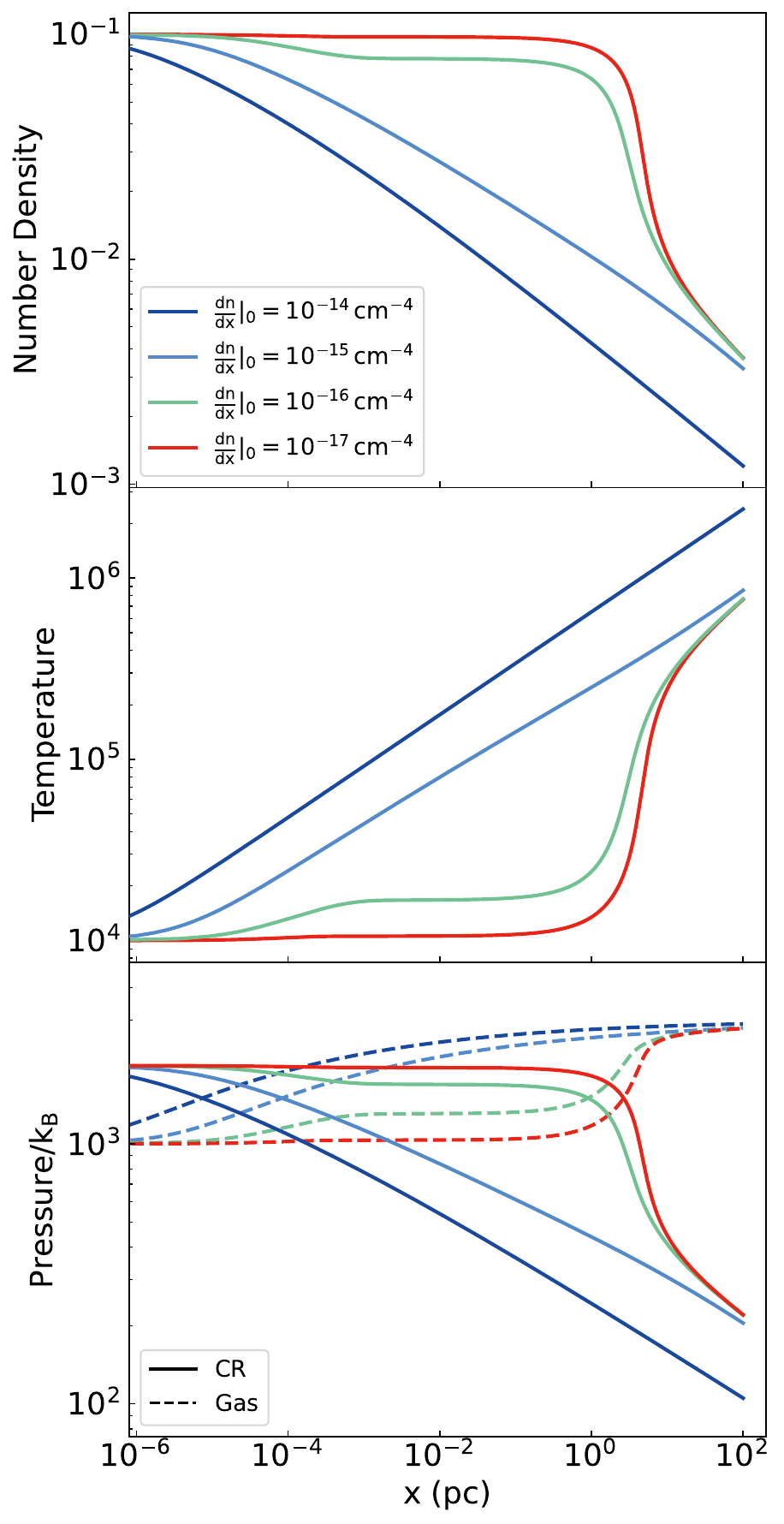}
\caption{Number density (top), temperature (middle), and pressure (bottom) profiles for a static front system, where the velocity $v=0$ and only thermal conduction, cosmic ray heating, and radiative cooling interact. The profiles are differentiated by the initial gradients of number density, $\frac{dn}{dx}\big|_0$, across a range from $10^{-14}$ to $10^{-17}$ cm$^{-4}$. We emphasize that except for the initial gradients, all profiles originate from the same initial conditions. The top panel shows the number density profiles, where steeper initial gradients lead to a more rapid decline in number density from the origin. In contrast, the less steep initial gradients see a substantial and sharp decrease around $\sim 1$ pc, indicating a delayed transition. The middle panel shows the corresponding temperature profiles. The different profiles underscore the significant impact of the initial gradient on the structure of the front. The bottom panel illustrates the pressure profiles; while the initial cosmic ray pressure is twice that of the gas pressure, it gradually decreases as the density drops.
}\label{fig:profile_alpha=3_static}
\end{figure}

In this section, we first examine the simpler static front case (where $v = 0$) to isolate and better understand the interaction between thermal conduction, cosmic ray heating, and radiative cooling. We numerically solve the energy equation (\ref{eq:gas3})
\begin{equation} \label{eq:static}
\pp{}{x}\left(-\kappa_T\pp{T}{x}\right)=-\rho\mathcal{L}-v_A\pp{P_c}{x}.
\end{equation} 
using forward integration and show the resulting number density, temperature, and pressure profiles as a function of distance in Figure~\ref{fig:profile_alpha=3_static}. Following \cite{Wiener2017}, we integrate to 100 pc. These profiles are distinguished by their initial gradients of number density, $\frac{dn}{dx}\big|_0$, with values ranging from $10^{-14}$ to $10^{-17}$ cm$^{-4}$. All the models are initialized with cosmic ray pressure dominant ($P_{c0}/P_{g0}=2$), and all become gas pressure dominated at large distances, typically by about an order of magnitude. This is due to the polytropic relationship between $P_c$ and $\rho$, whereas the decrease in $\rho$ is compensated by an increase in $T$.
As shown in Figure~\ref{fig:profile_alpha=3_static}, steeper initial gradients (e.g., $10^{-14}$cm$^{-4}$, $10^{-15}$cm$^{-4}$), which imply stronger influences from thermal conduction and cosmic ray heating, resulting in a noticeable and immediate decrease in number density from the origin. In contrast, profiles with less steep initial gradients (e.g., $\frac{dn}{dx}\big|_0 \leq 10^{-16}$ cm$^{-4}$), only show a significant decrease in number density at $\sim 1$ pc, indicating a delayed transition. Regardless of the initial gradient, the number density always decreases to of order $10^{-3}$ cm$^{-3}$ at $x=100$ pc, with the steepest gradient reaching $\sim 1\times10^{-3}$ cm$^{-3}$ and the least steep gradient dropping to $\sim 3\times10^{-3}$ cm$^{-3}$. The same behavior is observed in the temperature profiles in the middle panel of Figure~\ref{fig:profile_alpha=3_static}. Lower $\frac{dn}{dx}\big|_0$ values give profiles that overlap the $\frac{dn}{dx}\big|_0 = 10^{-17}$ cm$^{-4}$ one; we display one profile here for clarity. The bottom panel presents the pressure profiles for all four $\frac{dn}{dx}\big|_0 = 10^{-14}$ cm$^{-4}$ to $10^{-17}$ cm$^{-4}$ cases.

\begin{figure*}[!htb]
\centering
\includegraphics[width=\textwidth]{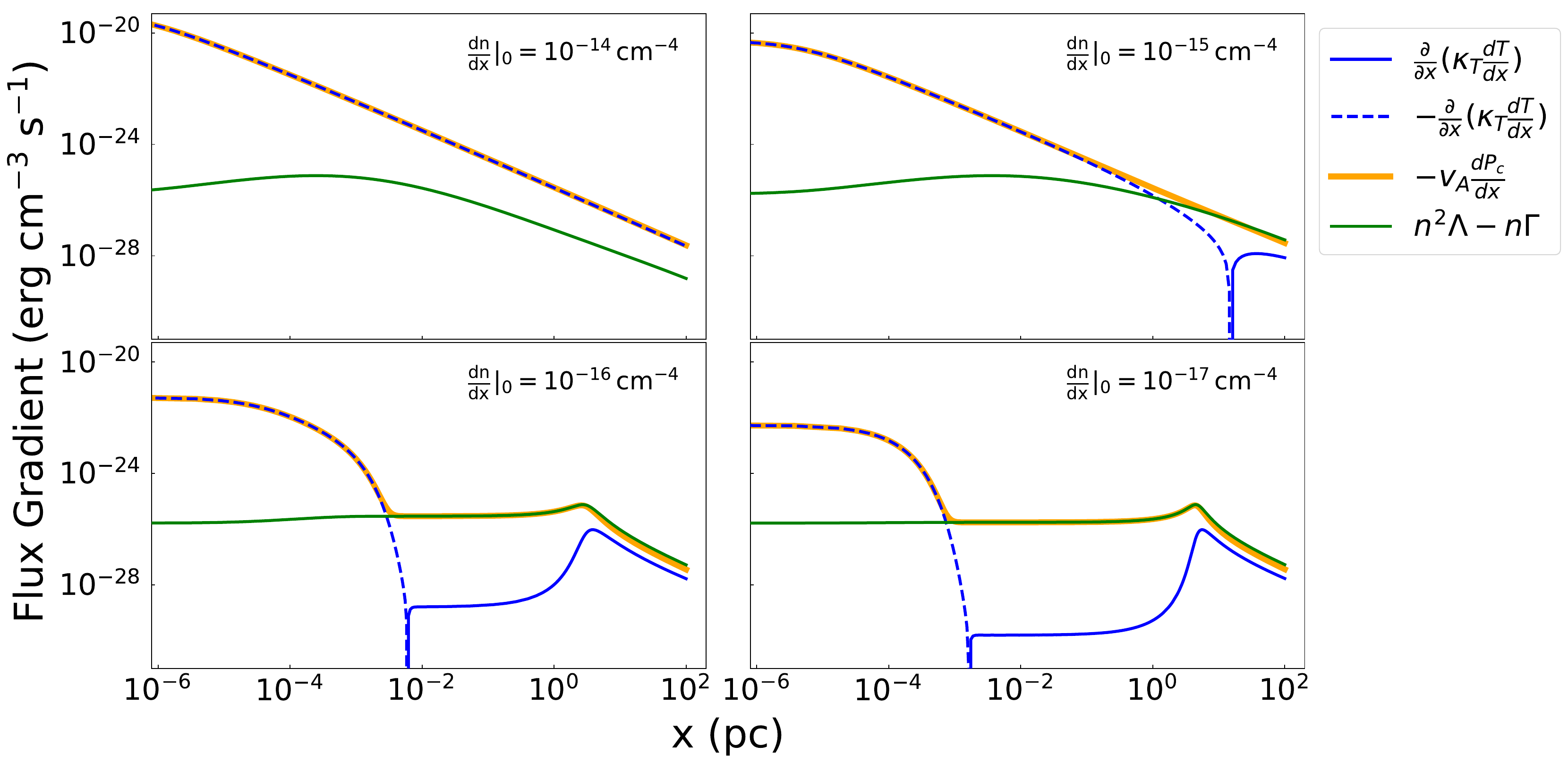}
\caption{Contributions of each flux component as a function of distance in the static case, in solid and dashed lines in erg cm$^{-3}$s$^{-1}$. This figure delineates the varying dynamics between different flux components under different initial number density gradients. For smaller initial number density gradients, cosmic ray heating first is balanced by conductive cooling, until it decreases to the level of radiative cooling, which is when conduction becomes less dominant heating term. For larger initial gradients, the transition from conductive cooling to heating occurs over a more extended distance (beyond 100 pc). The interplay between these components is critical to understanding the thermal dynamics of the system. }\label{fig:component_alpha=3_static}
\end{figure*}

In Figure~\ref{fig:component_alpha=3_static}, we show the contributions of each flux component as a function of distance. In the steepest initial gradient case, cosmic ray heating is consistently offset by conductive cooling. When $\frac{dn}{dx}\big|_0 \leq 10^{-15}$, after cosmic ray heating decreases to be smaller than radiative cooling, radiative cooling is then countered by CR heating and conductive heating. This balance continues all the way to the boundary at $x=100$ pc. It is not surprising that all these models require conductive cooling near the cloud, since the gradient required to balance cosmic ray heating with radiative cooling is of order 10$^{-21}$ cm$^{-4}$, much smaller than any of the models presented here.

Comparing our static front results with previous work highlights the influence of cosmic rays on the energy balance. In \citet{Cowie1977}, without cosmic rays, conductive heating balances radiative cooling at the front. In contrast, when cosmic rays are included, as in our model and \citet{Wiener2017}, cosmic ray heating becomes the dominant process balancing radiative cooling. Nonetheless, thermal conduction remains significant, as strong cosmic ray heating requires conductive cooling to dissipate the excess energy. Our result highlights the crucial role of thermal conduction in maintaining thermal equilibrium within the system.

\subsection{Evaporation Front} \label{sec:results:evaporation}
In this section, we incorporate an evaporative flow into the energy equation:
\begin{equation} \label{eq:evaporation}
\pp{}{x}\left(\frac{1}{2} \rho v^2 v + \frac{\gamma_g}{\gamma_{g}-1} P_g v-\kappa_T\pp{T}{x}\right)=-\rho\mathcal{L}-v_A\pp{P_c}{x}.
\end{equation} 

\begin{figure}[!htb]
\centering
\includegraphics[width=\columnwidth]{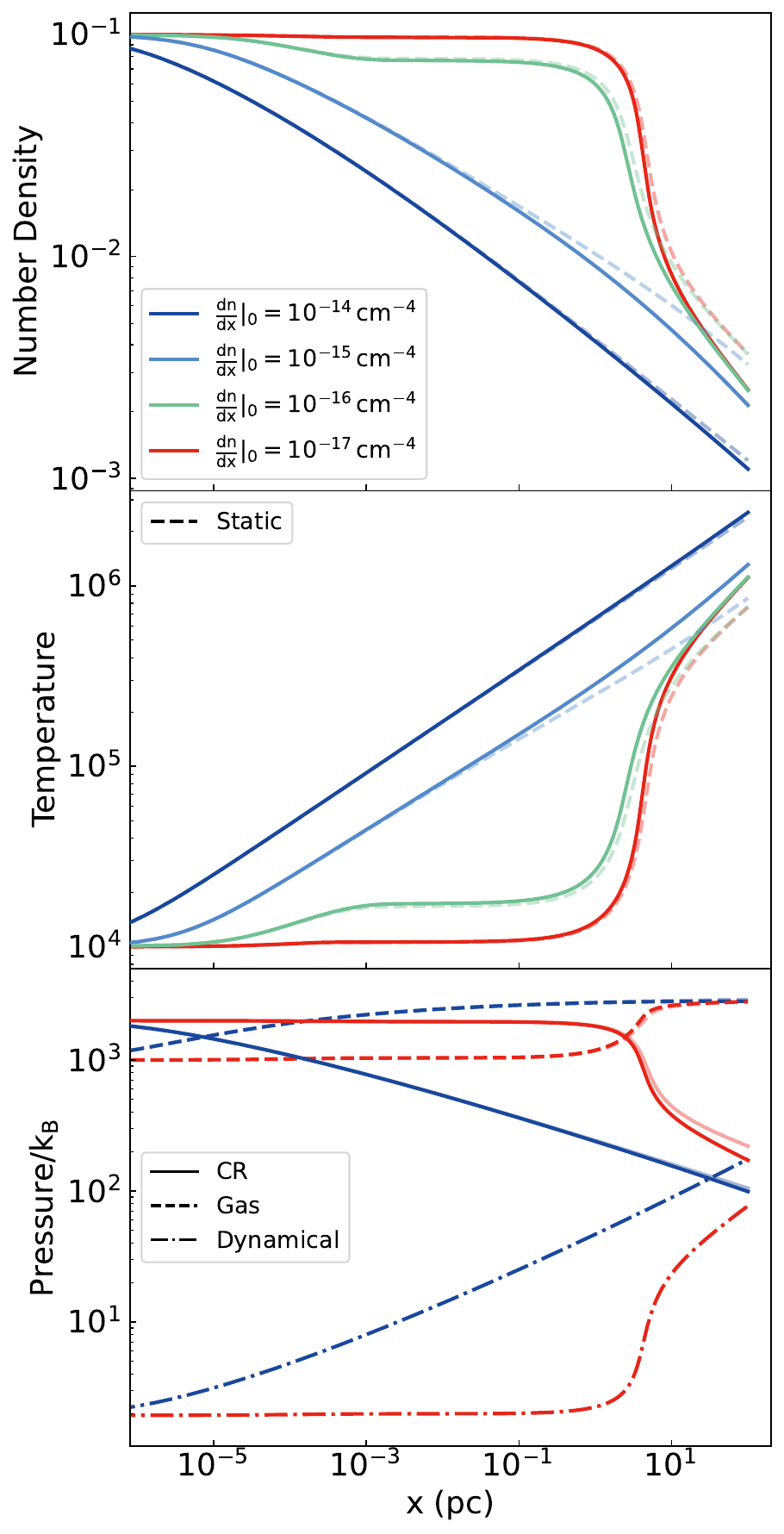}
\caption{Number density, temperature, and pressure profiles in the fiducial, evaporative front case. The dashed, semi-transparent lines show the flux magnitude for the static scenario. The top panel shows the number density profiles, where steeper initial gradients lead to a more rapid decline in number density from the origin. In contrast, the less steep initial gradients see a substantial and sharp decrease around $\sim 1$ pc, indicating a delayed transition. The middle panel shows the corresponding temperature profiles. The different profiles underscore the significant impact of the initial gradient on the structure of the front. The bottom panel illustrates the pressure profiles; while the initial cosmic ray pressure is twice that of the gas pressure, it gradually decreases as the density drops.}\label{fig:profile_alpha=3_evaporation}
\end{figure}

We now numerically solve Equation~\ref{eq:evaporation} and present the resulting number density, temperature, and pressure profiles in Figure~\ref{fig:profile_alpha=3_evaporation}. For comparison, the static front cases are shown with dashed lines. Notably, the profiles of the evaporation fronts begin to deviate from those of the static case beyond 0.1 pc, with these deviations amplifying with increasing distance. Specifically, the number density in the evaporation front scenario drops more significantly compared to the static case, which corresponds to a greater increase in temperature.  The pressure profiles show that as we move away from the origin, the cosmic ray (CR) and gas pressure profiles evolve similarly to the static front case. However, after the point where $P_c = P_g$, the difference between $P_c$ and $P_g$ becomes greater compared to the static front scenario, indicating a sharper transition between the roles of CR pressure and gas pressure. The presence of flow introduces dynamical pressure, which is absent in the static case. In the evaporation scenario, the pressure profile shows a continuous decrease in the sum of CR and gas pressures, accompanied by an increase in dynamical pressure. However, the dynamical pressure remains relatively small, accounting for only about 3\% of the total pressure at $x=100$ pc, emphasizing the dominant role of CR and gas pressures in shaping the dynamics of the layer.

\begin{figure*}[!htb]
\centering
\includegraphics[width=\textwidth]{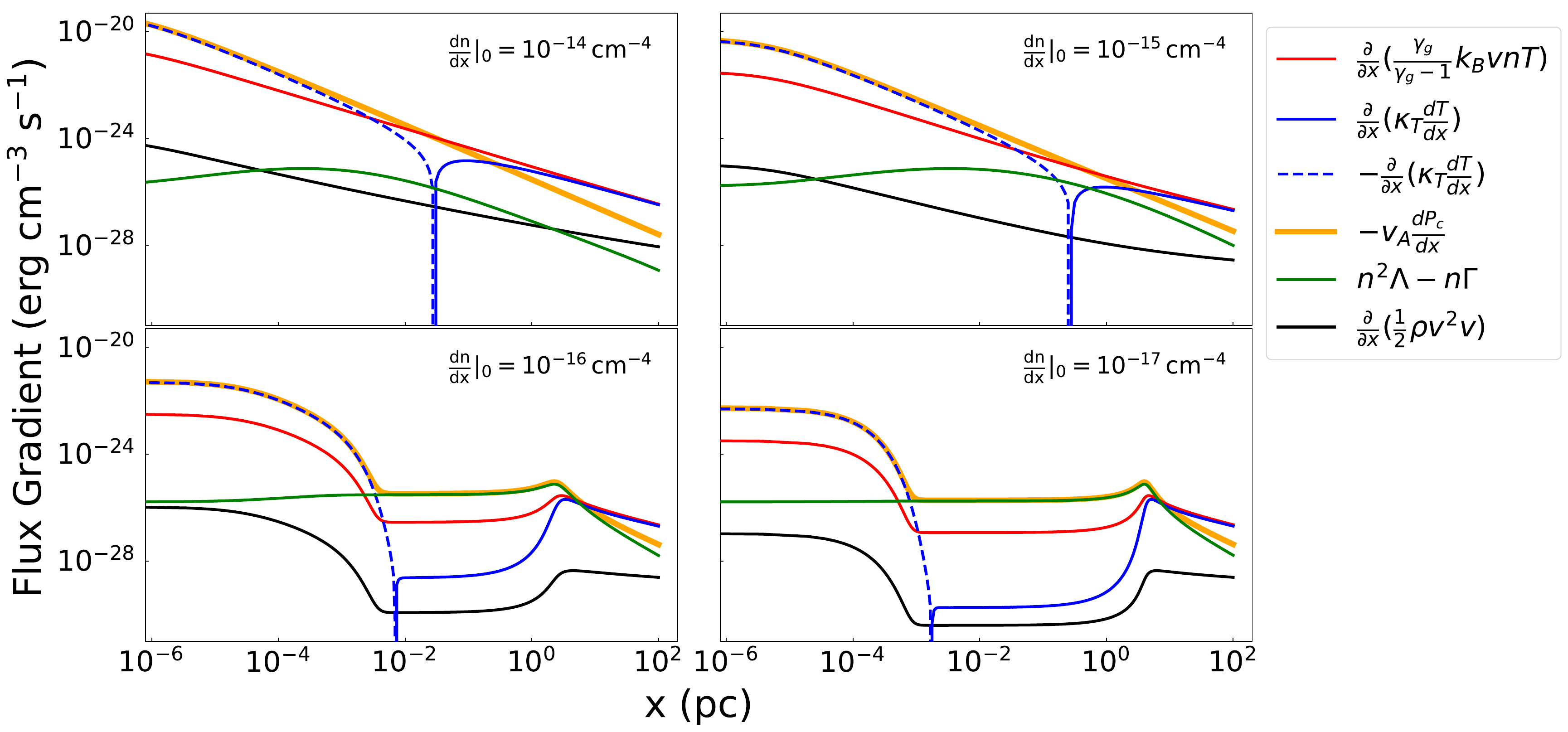}
\caption{Contributions of each flux component as a function of distance in the fiducial case. The thin, semi-transparent lines show the flux magnitude for the static scenario. For initial gradients less steep than $10^{-16}$ cm$^{-4}$, three distinct phases are identified: close to the origin, cosmic ray heating dominates and is counterbalanced by conductive cooling; at an intermediate distance, a transition to conductive heating occurs as cosmic ray heating balances radiative cooling; and beyond, where conductive heating dominates against declining radiative cooling. For steeper initial gradients ($\frac{dn}{dx}\big|_0 \geq 10^{-15}$ cm$^{-4}$), CR heating is so strong that radiative cooling cannot balance it, precluding the formation of regions with flat number density and temperature profiles.
}\label{fig:component_alpha=3_evaporation}
\end{figure*}

We also present the contributions of each flux component as a function of distance in Figure~\ref{fig:component_alpha=3_evaporation}. For less steep initial gradients (e.g., $\frac{dn}{dx}\big|_0 \leq 10^{-16}$ cm$^{-4}$), we identify three distinct regions. Near the origin, cosmic ray (CR) heating is the dominant process and is balanced by conductive cooling. Further from the origin, there is a transition region where CR heating decreases to levels comparable to radiative cooling, causing conduction to shift from cooling to heating. This transition indicates a regime where CR heating is balanced by radiative cooling. Beyond this point, as radiative cooling reaches its maximum and subsequently declines, conductive heating and enthalpy flux cooling become the dominant processes, similar to the Cowie \& McKee evaporation fronts. For steeper initial gradients (e.g., $\frac{dn}{dx}\big|_0 \geq 10^{-15}$ cm$^{-4}$), the dynamics change significantly. Strong CR heating prevents radiative cooling from fully balancing the thermal energy input, leading to the absence of a region where radiative cooling balances CR heating and thereby precluding the formation of regions with flat number density and temperature profiles. These results underscore the complex interplay between CR heating, radiative cooling, and conduction in regulating the thermal dynamics of the system.

\section{Varying Physical Parameters}\label{sec:background} 
In this section, we examine the effects of changing physical parameters, specifically the magnetic field strength and the ratio between cosmic ray pressure and gas pressure at the $x=0$ boundary, denoted as $\alpha \equiv P_{c,0}/P_{g,0} + 1$. The magnetic field strength appears only as a multiplier in the cosmic ray heating term. Meanwhile, the ratio $\alpha$ sets the initial cosmic ray pressure and also determines the system's total pressure, with a higher $\alpha$ indicating increased initial CR pressure and total pressure. We investigate how variations in these parameters impact the dynamics and structure of the layer.

\subsection{Cosmic Ray and Total Pressure}\label{ss:pc}

\begin{figure*}[!htb]
\centering
\includegraphics[width=\textwidth]{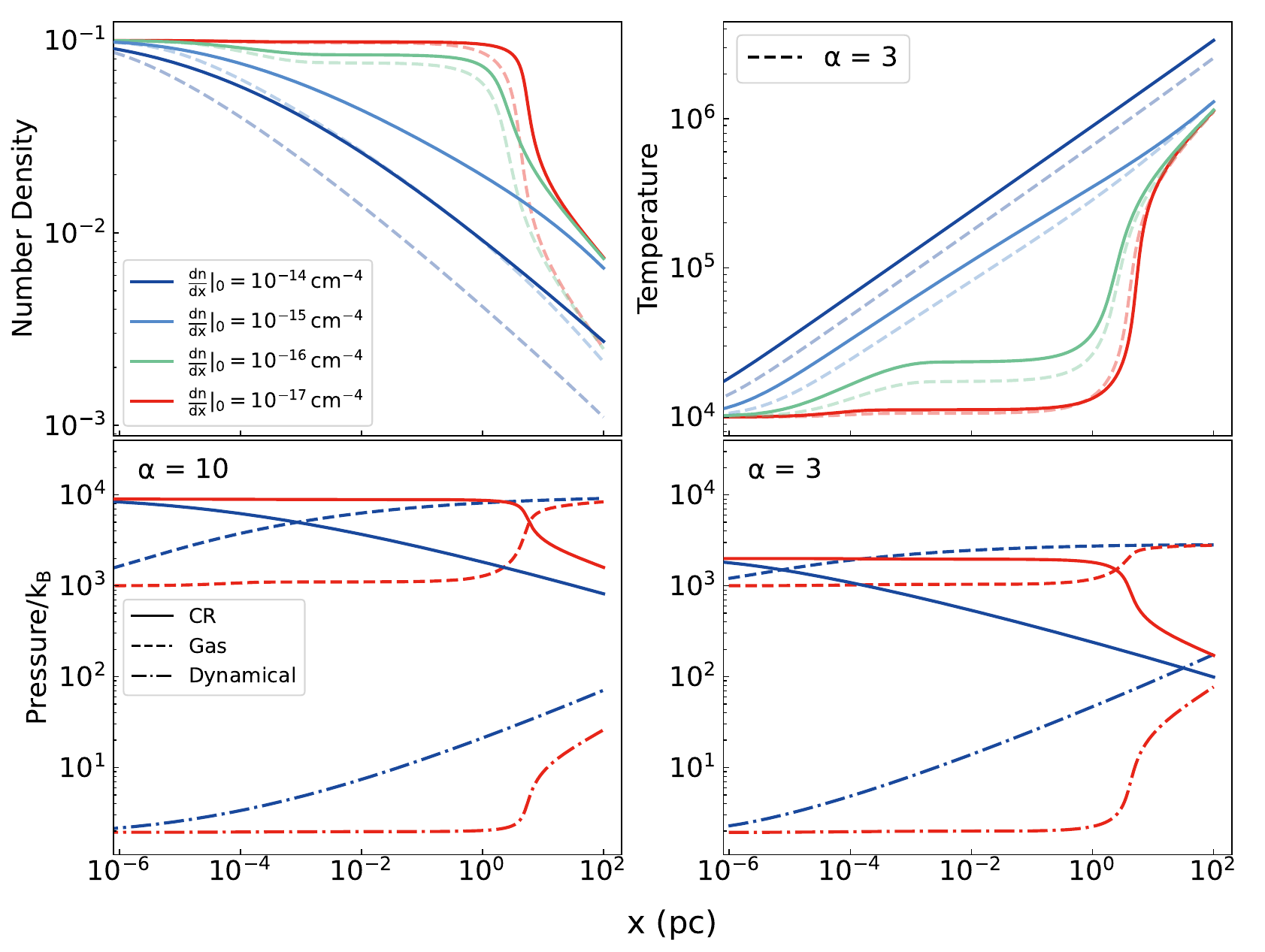}
\caption{Profiles for number density (top left), temperature (top right), and pressure (bottom left and right) for the evaporation scenario with a high cosmic ray pressure to gas pressure ratio at the $x=0$ boundary ($\alpha = 10$). In the top panels, the solid lines represent $\alpha = 10$, while the dashed semi-transparent lines depict the profiles for $\alpha = 3$. The top panels illustrate the impact of $\alpha$ on the number density and temperature profiles, showing that the number density reaches only $\sim 10^{-2}$ cm$^{-3}$ at $x=100$ pc for $\frac{dn}{dx}\big|_0 \leq 10^{-15}$ cm$^{-4}$, compared to $\sim 10^{-3}$ cm$^{-3}$ in the fiducial case. The bottom panels show the pressure profiles for the highest and lowest initial density gradients in blue and red, respectively, indicating that the transition from cosmic ray pressure dominated to gas pressure dominated persists at higher $\alpha$, while dynamical pressure constitutes only 0.8\% of the total pressure at the $x=100$ pc boundary, in contrast to the $\alpha = 3$ case where it is comparable to CR pressure.}\label{fig:profile_alpha_10_evaporation}
\end{figure*}

\begin{figure*}[!htb]
\centering
\includegraphics[width=\textwidth]{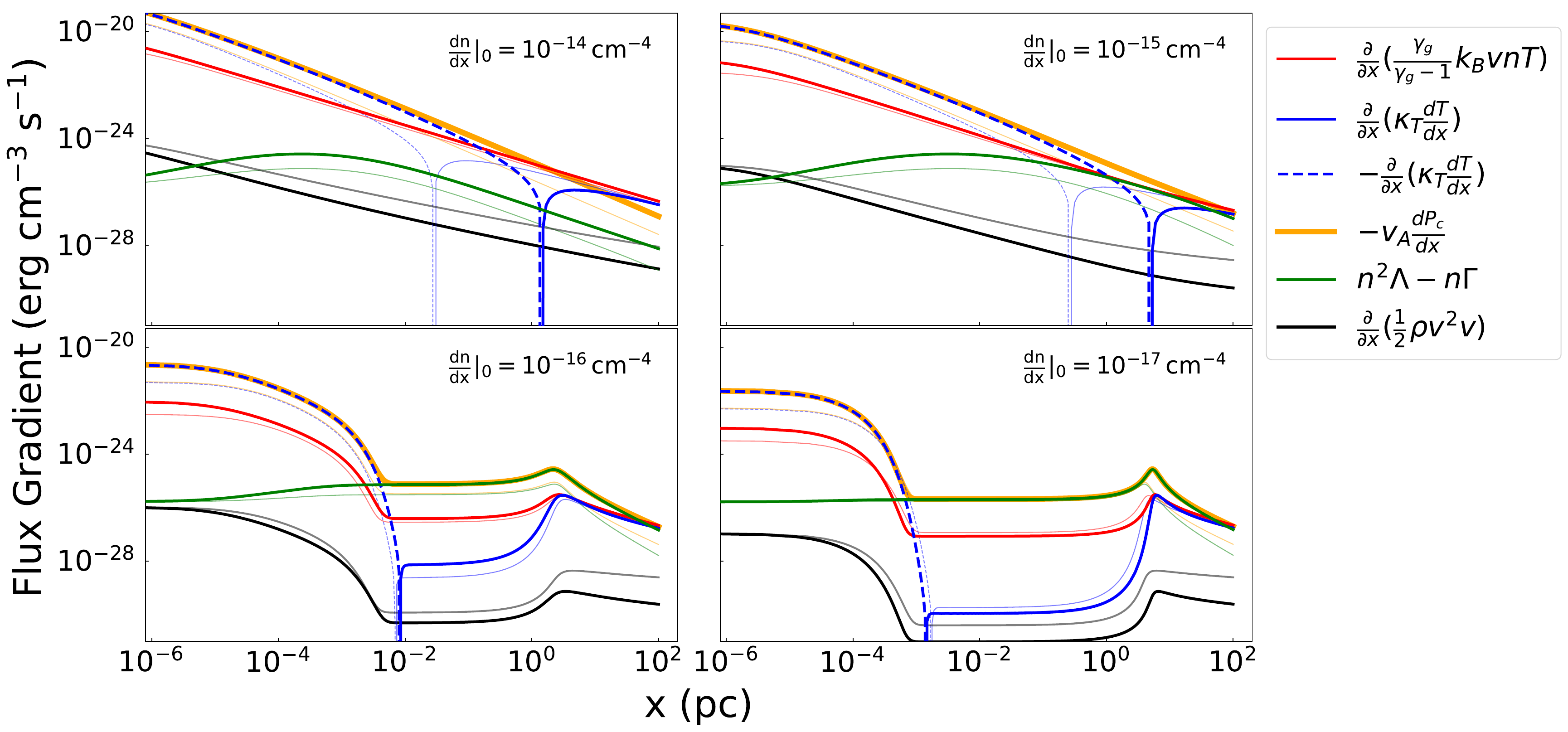}
\caption{Flux component contributions in the evaporation scenario with $\alpha = 10$, plotted over distance. Thin, semi-transparent lines indicate the $\alpha = 3$ fiducial case for comparison. The top two panels illustrate that for $\frac{dn}{dx}\big|_0 \geq 10^{-15}$ cm$^{-4}$, stronger CR heating at the origin extends the distance needed for conduction to change from cooling to heating. The bottom two panels show that for $\frac{dn}{dx}\big|_0 \leq 10^{-16}$ cm$^{-4}$, the number density gradient, and hence the temperature gradient, are steeper than their $\alpha = 3$ counterparts. The component plots demonstrate the impact of CR heating on the front structure and the spatial extent of physical processes.}\label{fig:component_alpha_10_evaporation}
\end{figure*}

In this subsection, we discuss the results for $\alpha = 10$. With this higher value, at the $x=100$ pc boundary, the number density does not drop as low, reaching only $\sim 10^{-2}$ cm$^{-3}$ for $\frac{dn}{dx}\big|_0 \leq 10^{-15}$ cm$^{-4}$, compared to $\sim 10^{-3}$ cm$^{-3}$ in the fiducial case. The impact of $\alpha$ on the temperature profile is smaller, however. Additionally, the role of dynamical pressure across the layer becomes much less significant. At the $x=100$ pc boundary, dynamical pressure constitutes only 0.8\% of the total pressure, in contrast to the $\alpha = 3$ case, where dynamical pressure at the same boundary is comparable to CR pressure.

Examining the component plot (Figure~\ref{fig:component_alpha_10_evaporation}) reveals the impact of $\alpha$. For $\frac{dn}{dx}\big|_0 \geq 10^{-15}$ cm$^{-4}$, stronger CR heating at the origin extends the distance needed for conduction to change from cooling to heating. This difference is illustrated in the top two panels of Figure~\ref{fig:component_alpha_10_evaporation}. We also observe that for $\frac{dn}{dx}\big|_0 \leq 10^{-16}$ cm$^{-4}$, the number density gradient, and hence the temperature gradient, are steeper than their $\alpha = 3$ counterparts (bottom panels of Figure~\ref{fig:component_alpha_10_evaporation}). Additionally, after the peak in radiative cooling, the $\alpha = 10$ scenarios require a longer distance for radiative cooling and CR heating to fall to levels similar to enthalpy flux and conduction. This extended distance is reflected in the more gradual changes in the number density and temperature profiles in this region. The component plots demonstrate how the strength of CR heating at the origin impacts the system's thermal structure and the spatial extent of physical processes. These observations are similar to the $\alpha = 3$ case, where the presence of CR heating leads to the adjustment of other terms in the energy equations to counteract CR heating.

\subsection{Magnetic Field Strength}

\begin{figure*}[htb!]
\centering
\includegraphics[width=\textwidth]{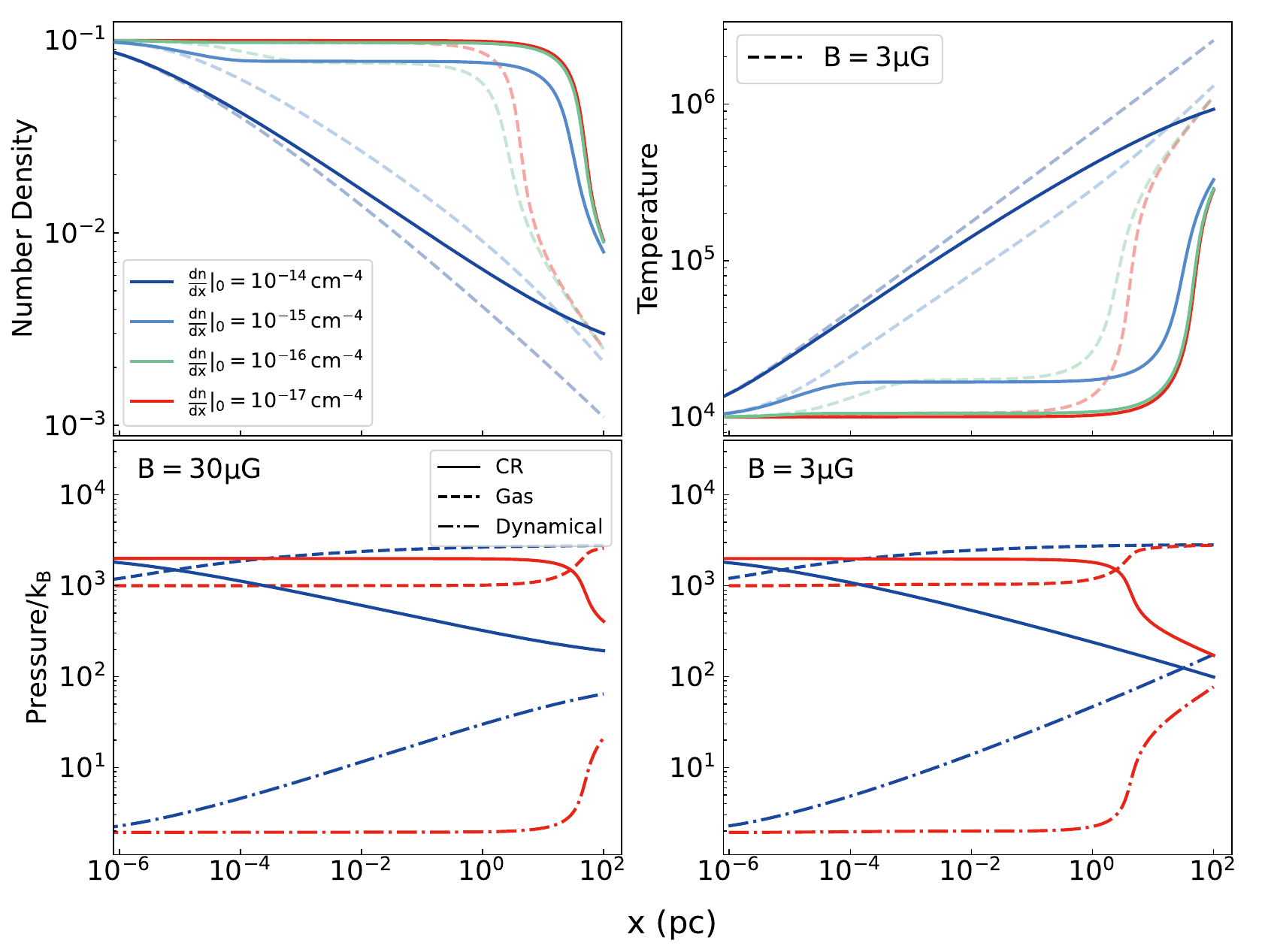}
\caption{Number density (top left), temperature (top right), and pressure (bottom) profiles for the evaporation scenario with a magnetic field strength of 30 $\mu$G and a cosmic ray to gas pressure ratio ($\alpha$) of 3. In the top panels, the solid lines represent the current scenario, while the thin lines show the profiles with a magnetic field strength of 3 $\mu$G. The profiles indicate that with a stronger magnetic field, dynamical pressure is reduced. Additionally, achieving similarly steep profiles as seen with a weaker magnetic field requires steeper initial number density gradients.}\label{fig:profile_B=30_alpha=3_evaporation}
\end{figure*}

We now investigate the impact of a heightened magnetic field strength, $B$, set to 30 $\mu$G. Note that when only $B$ is adjusted, the total pressure of the system remains the same; only the cosmic ray heating term increases by a factor of 10. With this stronger magnetic field, achieving number density and temperature profiles as steep as in previous cases requires steeper initial number density gradients. This change is clearly shown in Figure~\ref{fig:profile_B=30_alpha=3_evaporation}. A key difference between the $B = 30 \mu$G result and the $\alpha = 10$ result is that, with constant total pressure, the temperature does not reach $\sim 10^{6}$ K, as it does in the fiducial case and in the $\alpha = 10$ case. Similarly, the enhanced magnetic field reduces the relative importance of dynamical pressure compared to the trend in the fiducial case (see bottom panels of Figure~\ref{fig:profile_B=30_alpha=3_evaporation}).

\begin{figure*}[htb!]
\centering
\includegraphics[width=\textwidth]{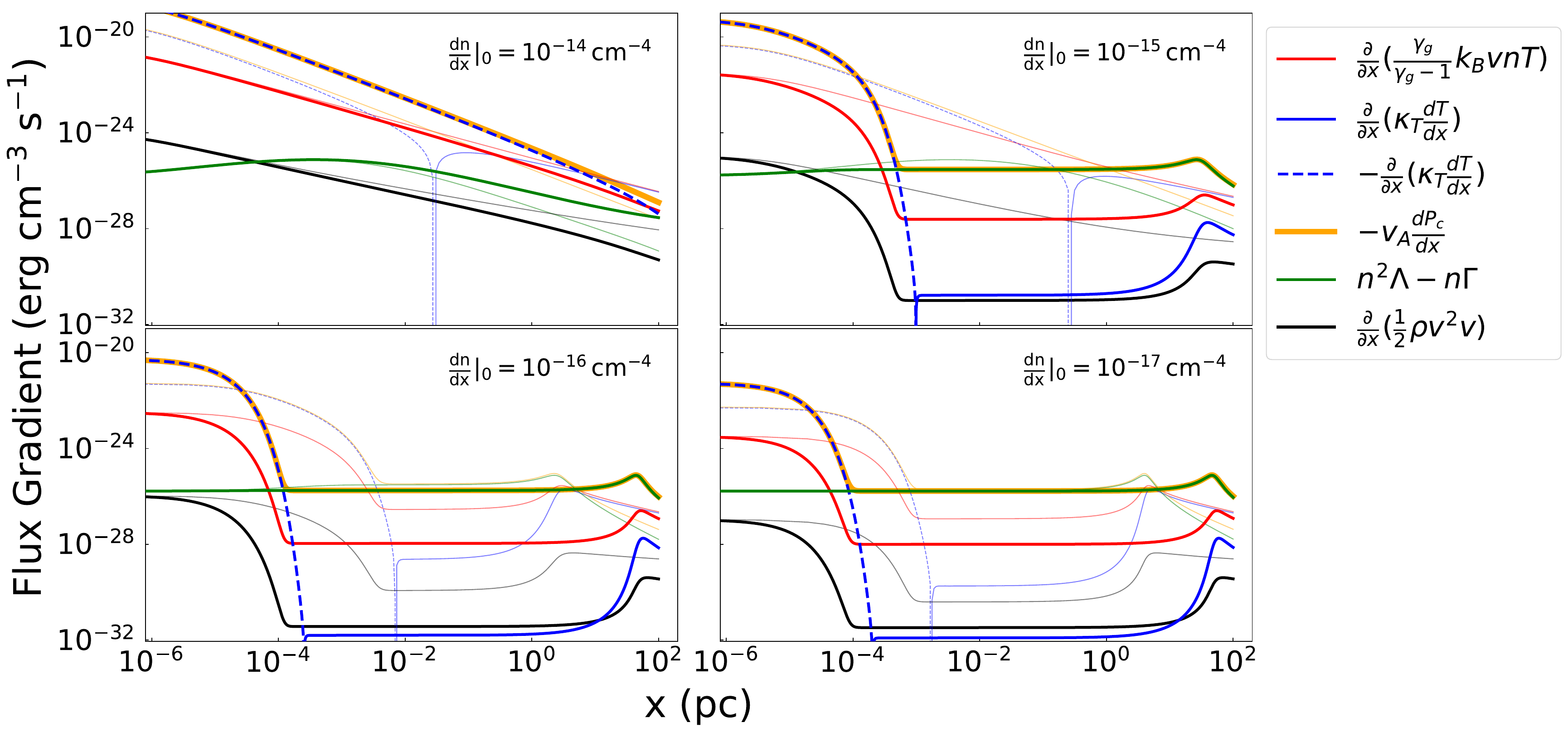}
\caption{Flux component contributions in the evaporation scenario with $\alpha = 3$ and a magnetic field strength of 30 $\mu$G. Solid lines represent the heightened magnetic field scenario, while thin, semi-transparent lines depict the flux contributions for a magnetic field strength of 3 $\mu$G. The plot shows a more rapid decline in CR heating, aligning more quickly with radiative cooling levels. Additionally, beyond the cooling function peak, CR heating and radiative cooling remain the dominant processes due to the smaller decline in number density.}\label{fig:component_B=30_alpha=3_evaporation}
\end{figure*}

Turning to Figure~\ref{fig:component_B=30_alpha=3_evaporation}, we see that a tenfold increase in the magnetic field strength causes CR heating to decrease more rapidly, aligning more quickly with radiative cooling levels. For all initial density gradients, by $x=100$ pc, the relatively smaller decrease in number density compared to the fiducial case means that enthalpy flux cooling and conductive heating are no longer the dominant processes. Instead, CR heating and radiative cooling become dominant at the $x=100$ pc boundary. This sustained dominance of CR heating and radiative cooling at larger scales is a direct outcome of the stronger magnetic field's influence on the system.

It may seem paradoxical that increasing $B$, and with it $v_A$, increases the heating rate but decreases $T$ far from the cloud. We can understand this by observing that as $B$ increases, so does the range over which cosmic ray heating is balanced by conductive cooling. This in turn requires that $T$
cannot increase too quickly. For example, it can readily be shown that if $T$ has a power law dependence on position, $T\propto x^p $, conductive cooling requires $p < 2/7$. This accounts for the relatively slow increase in $T$ in the high magnetic field models. 
\section{Ion Column Density Ratios}\label{sec:ratios}

We use the ionization fractions $X_i(T)$ from \citet{Gnat2007} and the relative abundances $A_X$ from \citet{Asplund2009} to compute the column densities of specific ions. The column density for each ion is defined by $N_X = A_X \int X_i(T) n \, \mathrm{d}x$, where $n$ is the number density, and the integral is across the front, up to a distance of 100 pc. We discuss this choice of integration boundary in Appendix~B. We then compare our calculated line ratios with measurements from high-velocity clouds (HVCs) observations in the Milky Way from \citet{Wakker2012}. Discrepancies between our theoretical line ratios and the observational data could highlight potential shortcomings in the current models.

We compute the column densities for \text{CIV}, \text{OVI}, and \text{SiIV}, and derive the diagnostic line ratios \text{SiIV}/\text{CIV}, \text{CIV}/\text{OVI}, and \text{NV}/\text{OVI}. These ratios serve as key indicators for understanding the ionization processes and thermal structure within the interstellar medium. The \text{SiIV}/\text{CIV} ratio probes cooler, photoionized gas with temperatures in the range of $10^4 - 10^5$ K, primarily indicating regions where UV radiation dominates the ionization processes \citep{Gnat2007, Wakker2012}. In contrast, the \text{CIV}/\text{OVI} ratio spans the transition between photoionized and hotter, collisionally ionized gas, typically found at temperatures between $10^5 - 10^6$ K. This ratio is particularly effective in identifying multiphase structures within the ISM, often tracing shock-heated gas in more dynamic regions \citep{Wakker2012, Shull2012}. Finally, the \text{NV}/\text{OVI} ratio is sensitive to collisionally ionized gas in the warm-hot ionized medium (WHIM), making it a useful diagnostic for gas in intermediate ionization states and the presence of shock-heated or disturbed gas \citep{Savage2009, Richter2017}.

We show in the left panel of Figure~\ref{fig:line_ratios} our model-dependent $\text{SiIV}/\text{CIV}$ ratio versus $\text{CIV}/\text{OVI}$ ratio in logarithmic scale, while in the right panel we present $\text{NV}/\text{OVI}$ ratio versus $\text{CIV}/\text{OVI}$ ratio, overplotted with observational data from \citet{Wakker2012}. The black circular data points with error bars represent the observational measurements, and our model predictions are superimposed as colored geometric shapes. Additionally, we include a CR heating balancing radiative cooling model, as discussed in \citet{Wiener2017}, shown in orange.

Our results reveal that most of the models lie well outside the observational range in the $\text{SiIV}/\text{CIV}$ versus $\text{CIV}/\text{OVI}$ parameter space. The highest magnetic field ($B$) value models predict $\text{SiIV}/\text{CIV}$ ratios that are consistent with the observations; however, these models simultaneously predict $\text{CIV}/\text{OVI}$ ratios that are too high compared to the observational data. In contrast, models with lower magnetic field values produce $\text{SiIV}/\text{CIV}$ ratios that are significantly lower than those observed in the HVCs.

Similarly, in the right panel, the high $B$ value models systematically overestimate the $\text{NV}/\text{OVI}$ ratio, placing them outside the range of observational constraints. For the other models, those with the largest $\frac{dn}{dx}\big|_0$ underestimate the $\text{CIV}/\text{OVI}$ ratio, suggesting that the structure and dynamics of the gas in these models may be insufficient to fully capture the physical processes driving the ionization conditions. However, the  range of models that overlap the data plotted in (N V/O VI, C IV/O VI) space is quite large, suggesting that this is not a particularly sensitive diagnostic of conditions in the front, including the initial value of the density gradient.

A \textit{prima facie} interpretation of these results suggests that if a model lies within the observational range in the $\text{NV}/\text{OVI}$ versus $\text{CIV}/\text{OVI}$ parameter space but fails to reproduce the correct $\text{SiIV}/\text{CIV}$ ratio, it implies that the model is better at predicting the ionization profile for gas in the $10^5 - 10^6$ K temperature range than for cooler gas at $10^4 - 10^5$ K. This also indicates that the model more accurately captures conditions dominated by collisional ionization rather than photoionization. These discrepancies highlight potential limitations in the current model framework, particularly regarding the role of magnetic fields and cosmic ray heating in shaping the ionization state of the ISM and CGM. In fact, the line ratios are not well fit by previous models which do not include cosmic rays (see Figure 11 of \cite{Wakker2012}). Our $\alpha=1$ models, as shown in Figure~\ref{fig:line_ratios}, align with this finding.

It should not be overlooked, however, that some of our models do successfully reproduce both pairs of line ratios. As might be expected from the forgoing discussion, these are the models with intermediate field strength;  $B=20\mu G$ for $\alpha = 3$.  As we already noted, the fits are relatively insensitive to the initial density gradient $\frac{dn}{dx}\big|_0$, which sets the initial cosmic ray heating rate and which we regard as the least well-determined parameter in our models.

\begin{figure*}[htb!]
    \begin{minipage}{0.49\textwidth}
        \centering
        \includegraphics[width=0.99\textwidth]{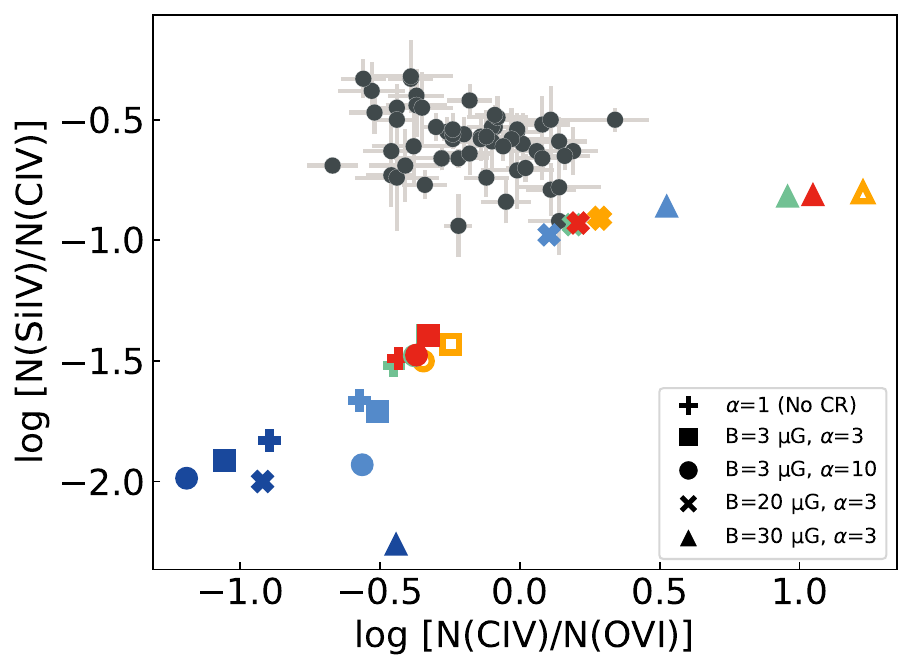}
    \end{minipage}\hfill
    \begin{minipage}{0.49\textwidth}
        \centering
        \includegraphics[width=0.99\textwidth]{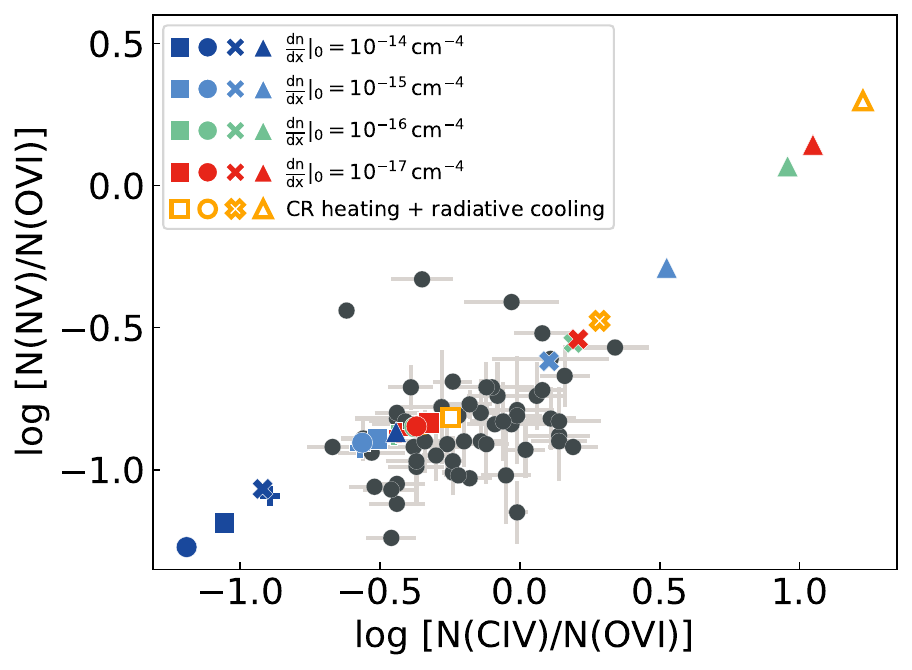}
\end{minipage}
\caption{Left panel: $\text{SiIV}/\text{CIV}$ versus $\text{CIV}/\text{OVI}$ ratios in logarithmic scale, compared with observational data from \citet{Wakker2012} (black circles with error bars). Right panel: $\text{NV}/\text{OVI}$ versus $\text{CIV}/\text{OVI}$ ratios, similarly overlaid with observational data. Colored shapes denote our model predictions, with variations in magnetic field strengths and cosmic ray parameters. The orange shapes represent models where CR heating balances radiative cooling, following \citet{Wiener2017}. Both panels reveal discrepancies between the models and observations, particularly in the $\text{SiIV}/\text{CIV}$ versus $\text{CIV}/\text{OVI}$ space, where no models fit within the observed range. High magnetic field models match $\text{SiIV}/\text{CIV}$ but overestimate $\text{CIV}/\text{OVI}$, while lower field models miss $\text{SiIV}/\text{CIV}$ but align with $\text{CIV}/\text{OVI}$ and $\text{NV}/\text{OVI}$. This suggests that in general, the models are more accurate at higher temperatures but struggle with cooler, photoionized conditions. Note that models with intermediate field strength, $B=20\mu$G, do overlap the date on both plots, although we make no claim to their uniqueness.}
\end{figure*}\label{fig:line_ratios}

\section{Summary and Discussion} \label{sec:discussion}

In this paper, we explored the evolution of the transition layer between warm ($10^4$ K) and hot ($10^6$ K) gas phases in the circumgalactic medium by integrating the energy equation (Equation~\ref{eq:gas3}) as an initial value problem. We showed that cosmic ray heating, thermal conduction, radiative cooling, enthalpy flux, and dynamical pressure collectively shape the temperature and density profiles of the transition layer. At large distances from the cloud, we found that conduction is the main heat source, and that it can be balanced by an outward enthalpy flux, as in the models of \cite{Cowie1977}.

We also found that the temperature and density profiles across the front are quite sensitive to the value we adopt for the initial density gradient. When the gradient is sufficiently small, radiative cooling suffices to balance cosmic ray heating as assumed in the front models of \cite{Wiener2017}. For sufficiently large gradients, cosmic ray heating is balanced by conductive cooling. In view of this sensitivity, we have to ask what controls this ratio in a natural environment. While addressing this question is beyond the scope of this paper, we speculate that the answer lies either in the front formation process or in the stability of the front to small perturbations. These possibilities will be examined in future work.

Our findings emphasize the dominant role that cosmic rays can play in regulating the structure of the transition layer between the warm and hot gas phases of the CGM. The addition of CR heating in these models dictates the thermal balance, with the other flux terms adjusting accordingly to balance CR heating. These effects are sensitive to the magnetic field strength and the ratio of CR to gas pressure ($\alpha$) at the warm gas boundary.

Our analysis of ion column density ratios shows that while our model predictions align with some observational data, and some represent successful fits, discrepancies persist. Only the highest magnetic field models reproduce the observed $\text{SiIV}/\text{CIV}$ ratios, but they overestimate the $\text{CIV}/\text{OVI}$ ratios. Conversely, other models underestimate the $\text{SiIV}/\text{CIV}$ ratios, although some fall within the observed range for $\text{CIV}/\text{OVI}$ and $\text{NV}/\text{OVI}$. The agreement with $\text{CIV}/\text{OVI}$ and $\text{NV}/\text{OVI}$, but not with $\text{SiIV}/\text{CIV}$, suggests that these models better represent ionization conditions at higher temperatures ($10^5 - 10^6$ K), where collisional ionization dominates, but struggle to capture the profiles in cooler, photoionized regions. These discrepancies indicate that while CR heating significantly influences the ionization structure of the CGM, additional processes such as turbulent mixing, shocks, or self-ionization may be necessary to fully explain the observed line ratios in lower temperature regimes. The need to include such effects would be particularly compelling if magnetic field strengths as high as 20$\mu$G (the magnetic field strength adopted in our most successful models)  could be ruled out.

In addition to the influence of cosmic rays, our study highlights the crucial role of thermal conduction in maintaining energy balance within the transition layers. Thermal conduction dissipates excess heat from cosmic ray interactions, preventing runaway heating and stabilizing the temperature profiles.

Overall, our findings reinforce the notion that the CGM is a complex and dynamic environment where a variety of physical processes, e.g. cosmic ray heating, radiative cooling, thermal conduction, and gas flows, interact to shape its structure and ionization state. Future work will focus on investigating the stability of these transition layers, particularly in the presence of cosmic ray pressure, magnetic fields, and thermal conduction. Studying the onset and development of instabilities within these fronts will be crucial for understanding their behavior.

\acknowledgments
We are happy to acknowledge useful discussions with Roark Habegger, Tsuyoshi Inoue, Andrey Kravtsov, Peng Oh, Brant Tan, Robert Benjamin, and Aaron Tran. This work was supported in part by the NSF grant AST 2010189 and in part by grant NSF PHY-2309135 and the Gordon and Betty Moore Foundation Grant No. 2919.02 to the Kavli Institute for Theoretical Physics (KITP). HZ thanks the Department of Astronomy at the University of Wisconsin-Madison for their hospitality. This manuscript has also been co-authored by Fermi Research Alliance, LLC under Contract No. DE-AC02-07CH11359 with the United States Department of Energy.

\appendix 

\section{Condensation Front}\label{appendix:condensation}

From the momentum conservation equation (Equation~\ref{eq:gas2}) in the steady-state case (i.e., assuming a zero time derivative), we have:
\begin{equation}
\frac{\partial}{\partial x} (\rho v^2 + P) = 0,
\end{equation}
where $P$ is defined as the total pressure, $P = P_g + P_c$, representing the sum of the gas pressure and the cosmic ray pressure.

Differentiating the equation, we have:
\begin{equation}
\frac{dP}{d\rho} = \frac{(\rho v)^2}{\rho^2}.
\end{equation}
Since $(\rho v)^2 / \rho^2$ is positive, this implies that $\frac{dP}{d\rho} > 0$. Therefore, a monotonically decreasing density $\rho$ corresponds to a monotonically decreasing pressure $P$.

This relationship implies that the flow cannot proceed from the hot medium to the cold medium. Specifically, in a scenario where the density is monotonically decreasing as the gas transitions from the cold to the hot medium, the corresponding pressure must also decrease. Consequently, gas flow is driven in the direction of decreasing pressure, which means it cannot move from the hot medium to the cold medium.

\section{Sensitivity of Line Ratios to Integration Cutoffs}

The line ratios we presented are sensitive to the cutoff point chosen for the integration over the transition layer. In Figure~\ref{fig:line_ratio_cutoffs}, we plot the line ratios ($\log (\text{CIV}/\text{OVI})$, $\log (\text{SiIV}/\text{CIV})$, and $\log (\text{NV}/\text{OVI})$) as functions of temperature where the integration is cut off, with the shaded regions representing the range of observational data. The vertical lines indicate the temperature at $x = 100$ pc for the two models: the red vertical line corresponds to the temperature for the model with $\frac{dn}{dx}\big|_0 = 10^{-17}$ cm$^{-4}$, and the blue vertical line corresponds to the model with $\frac{dn}{dx}\big|_0 = 10^{-14}$ cm$^{-4}$.

As seen in Figure~\ref{fig:line_ratio_cutoffs}, the choice of temperature cutoff has a significant, nontrivial effect on the calculated line ratios. \citet{Wiener2017} used a cutoff at $2 \times 10^6$ K, which aligns more closely with the converged values in the higher-temperature regime, compared to cutting off at 100 pc for the red model. However, while extending the integration to higher temperatures may yield more stable line ratios, the physical thickness of the transition layers becomes an important consideration. The models suggest that line ratios converge at higher temperatures, but the temperature at $x = 100$ pc remains well below this threshold for both models. This highlights a key issue: while we can theoretically integrate to very high temperatures, the question remains whether the transition layers are actually that thick in reality. Currently, we lack observational probes that can resolve the full extent of these layers. This uncertainty underscores the need for careful definition of the integration boundaries and comparison across models and observations.

\begin{figure*}[htb!]
\centering
\includegraphics[width=\columnwidth]{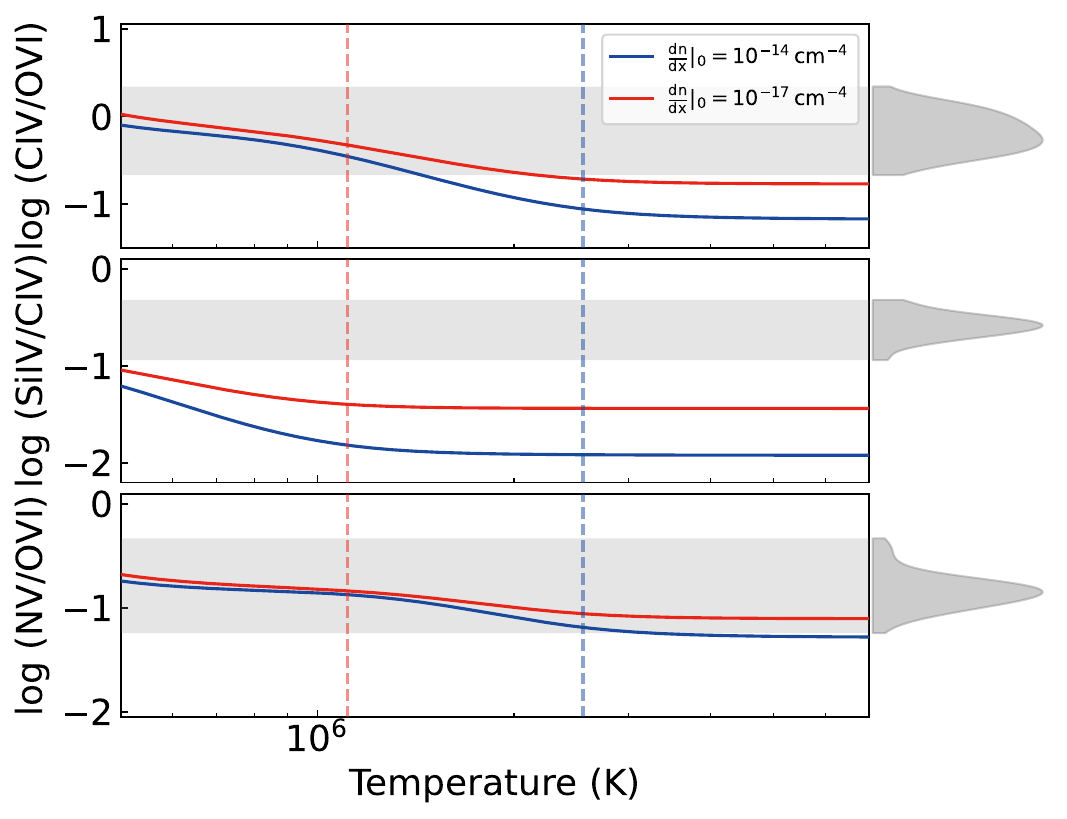}
\caption{Line ratios $\log (\text{CIV}/\text{OVI})$, $\log (\text{SiIV}/\text{CIV})$, and $\log (\text{NV}/\text{OVI})$ as a function of the temperature at which the integration is cut off. The shaded regions represent the range of observational data for each line ratio. The vertical red and blue lines correspond to the temperatures at $x = 100$ pc for the models with $\frac{dn}{dx}\big|_0 = 10^{-17}$ cm$^{-4}$ (red) and $\frac{dn}{dx}\big|_0 = 10^{-14}$ cm$^{-4}$ (blue), respectively. The plot illustrates the sensitivity of the calculated line ratios to the choice of integration cutoff.}\label{fig:line_ratio_cutoffs}
\label{fig:line_ratio_cutoffs}
\end{figure*}

\bibliographystyle{apj}
\bibliography{main}

\end{CJK*}
\end{document}